# The performances of R GPU implementations of the GMRES method

Bogdan Oancea
University of Bucharest
bogdan.oancea@faa.unibuc.ro

Richard Pospisil
Palacky University of Olomouc
richard.pospisil@upol.cz

Abstract

Although the performance of commodity computers has improved drastically with the introduction of multicore processors and GPU computing, the standard R distribution is still based on single-threaded model of computation, using only a small fraction of the computational power available now for most desktops and laptops. Modern statistical software packages rely on high performance implementations of the linear algebra routines there are at the core of several important leading edge statistical methods. In this paper we present a GPU implementation of the GMRES iterative method for solving linear systems. We compare the performance of this implementation with a pure single threaded version of the CPU. We also investigate the performance of our implementation using different GPU packages available now for R such as gmatrix, gputools or gpuR which are based on CUDA or OpenCL frameworks.

1. Introduction

Since the appearance of the first supercomputers in early '70s the nature of parallel computing has changed drastically. Besides the advancements in technology that allow building processors with a very high clock rate and a low specific dimension, today the commodity computers are based on multicore architecture that can run multiple tasks in parallel. Besides the classical multiheaded programming, another parallel computing paradigm called GP-GPU computing (general-purpose computing on graphics processing units), first experimented around 2000-2001 (see for example Larsen and McAllister, 2001), is widely used nowadays to speed up numerical intensive applications.

While numerical intensive parts of applications are handled traditionally by the CPUs, now GPUs have FLOP rates much higher than CPUs because GPUs are specialized for highly parallel intensive computations and they are designed with much more transistors allocated to data processing rather than control flow or data caching. This is shown in figure 1 where it can be noticed that the total surface of the device dedicated to ALUs is much higher in case of GPU than CPU.

The theoretical FLOP rates for GPUs and CPUs are presented in figure 2 and the memory bandwidth in figure 3. Current GPUs have FLOP rates at least 100 times greater than CPUs and the memory bandwidth is at least 10 times greater for GPU memories than for the main memory of the computer. These figures indicate that GPUs are ideal candidates to dispatch the numerical parts of applications.

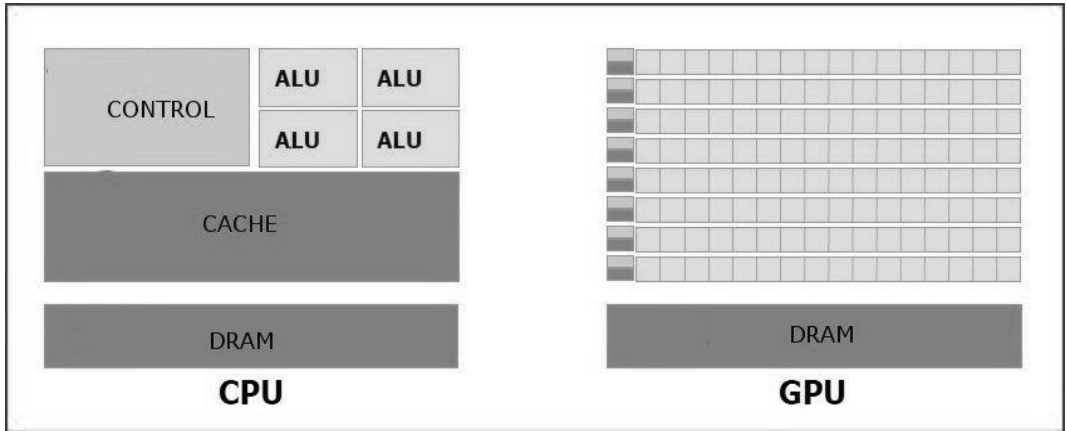

Figure 1. A comparison between CPU and GPU structure[1]

Statistical software relies heavily on numerical computations and can take advantage of the new paradigm of GPU computing. In this paper we will evaluate the performances of an R implementation of a numerical algorithm widely used in linear algebra libraries, namely the GMRES method to solve linear systems (Saad and Schultz, 1986). The rest of the paper is structured like this: in section 2 we briefly present the GPU computing frameworks currently available for scientific computing, in section 3 we present the GMRES algorithm, in section 4 we present some R packages that implements the GP-GPU computing model and discuss three R implementations that uses GPU computing, comparing their performances with the serial implementations. Section 5 concludes our paper.

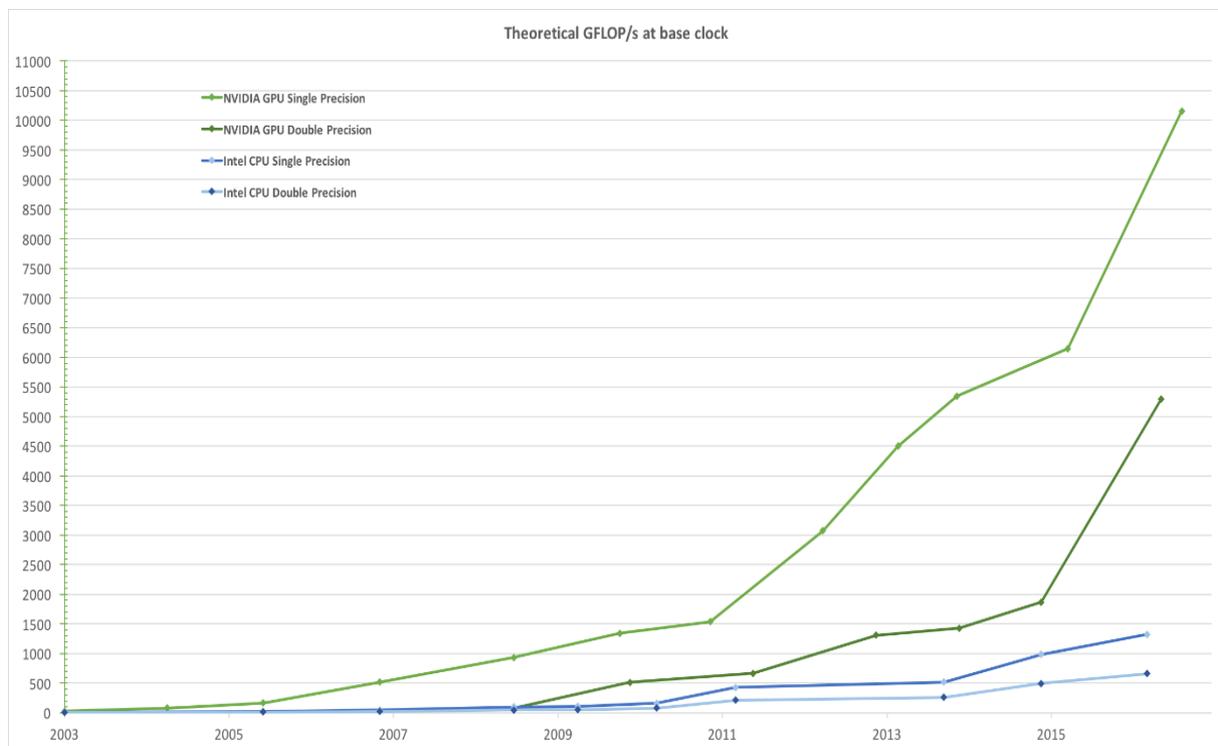

Figure 2. A comparison of the theoretical FLOP rates of typical CPUs and GPUs[2]

---

[1] http://docs.nvidia.com/cuda/cuda-c-programming-guide/
[2] http://docs.nvidia.com/cuda/cuda-c-programming-guide/

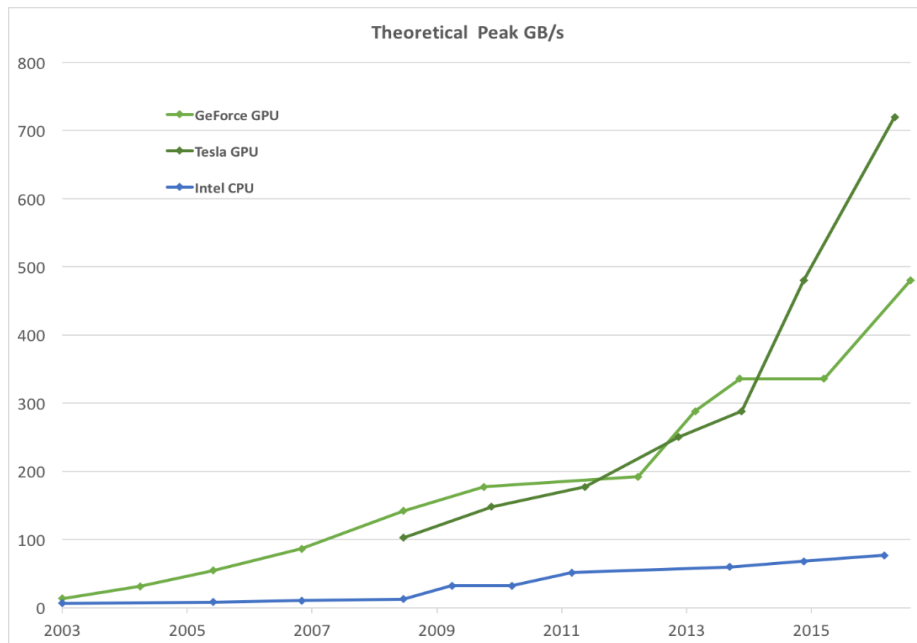
Figure 3. A comparison of the memory bandwidth for CPU and GPU[3]

2. GPU computing frameworks

GPUs are designed to solve general problems that can be formulated as data-parallel computations: the same stream of instructions is executed in parallel on many (different) data elements with a high ratio between arithmetic operations and memory accesses, like the SIMD approach of the parallel computers taxonomy. Currently there are several frameworks that implement GP-GPU model:

- CUDA (Compute Unified Device Architecture) – first introduced in 2006 by NVIDIA, is a general purpose parallel programming architecture that uses the parallel compute engines in NVIDIA GPUs to solve numerical intensive problems in a more efficient way than a CPU does (NVIDIA Corporation, 2007);
- OpenCL (Open Computing Language) - is a framework for writing programs that execute across heterogeneous platforms consisting of CPUs, GPUs, digital signal processors (DSPs), field-programmable gate arrays (FPGAs) and other hardware accelerators (OpenCL Specification Version 1.0, 2008);
- OpenACC - (Open Accelerators) is a programming standard for parallel computing developed by Cray, CAPS, NVIDIA and PGI. The standard is designed to simplify parallel programming of heterogeneous CPU/GPU systems – it works on NVIDIA, AMD and Intel accelerators (OpenAcc-Standard.org, 2017);

In this paper we used CUDA framework and the R packages that allows access to the CUDA API. CUDA parallel programming framework works with three important abstractions (Oancea et al., 2012a):
- a hierarchy of thread groups;
- shared memories;
- a barrier synchronization accessible by the programmer;

---

[3] http://docs.nvidia.com/cuda/cuda-c-programming-guide/

CUDA parallel programming paradigm requires programmers to (NVIDIA Corporation, 2017):
- partition the problem into coarse tasks that can be run in parallel by blocks of threads;
- divide each task into finer groups of instructions that can be executed cooperatively in parallel by the threads within a block;

The main programming language for CUDA is C but nowadays there are bindings to other languages too: Java, Fortran, Common Lisp, Python, R. While C libraries that uses CUDA for numerical computing are mature now, the bindings to other languages are only at the beginning. In (Oancea et al., 2012b) we investigated the performance of a Java library that uses CUDA for numerical computations and found that one can achieve an important speedup compared to the classical (serial) version of the respective algorithm while in (Oancea and Andrei, 2013) we investigated a hybrid approach CPU-GPU. In this paper we will test if a linear algebra algorithm implemented in R is capable to obtain a significant speedup compared with a serial R implementation.

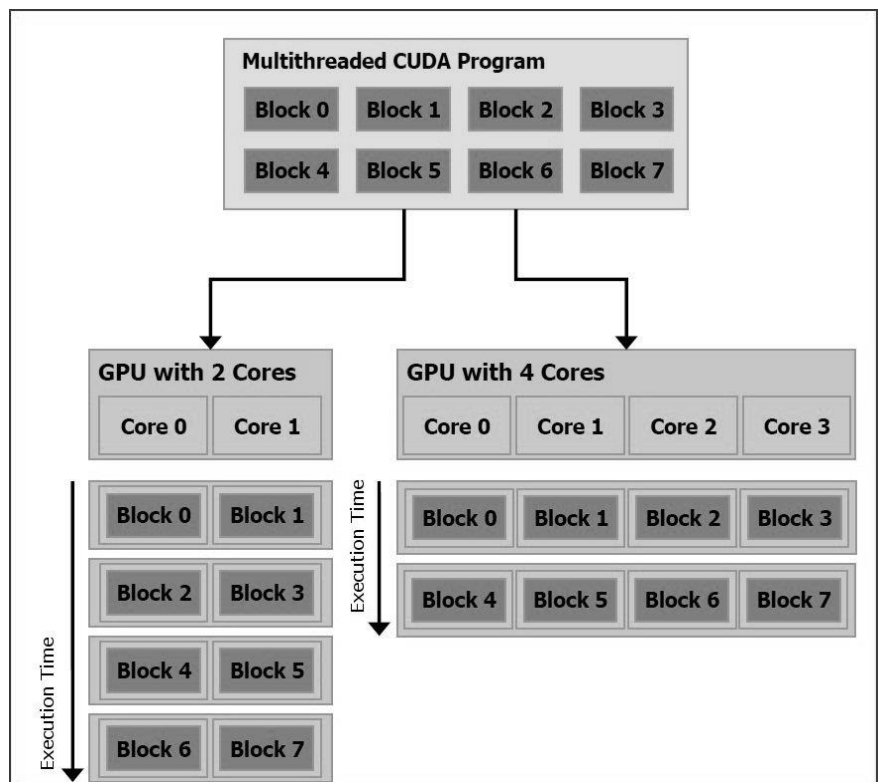

Figure 4. CUDA paradigm of GPU computing[4]

3. GMRES method

One question may arise: why GMRES? When using GPU for numerical computations the biggest speedups are obtained by algorithms that use level 3 BLAS operations (such as matrix-matrix multiplication). GMRES uses only level 1 and level 2 operations (vector updates and matrix-vector products) and we wanted to test if the available GPU R packages can obtain an important advantage (speedup) over classical numerical algorithms even for methods that are not easily parallelizable.

---
[4] http://docs.nvidia.com/cuda/cuda-c-programming-guide/

The GMRES (Generalized Minimum RESidual) was first proposed in 1986 as a Krylov subspace method for nonsymmetric systems (Saad and Schultz, 1986). This method approximates the solution by a vector in a Krylov subspace with minimal residual – this vector is build using the so-called Arnoldi iteration.

The problem to be solved is $Ax = b$. The n-th Krylov subspace for this problem is: $K_n = K_n(A, b) = span\{b, Ab, A^2b, \ldots, A^{n-1}b\}$. GMRES approximates the exact solution of the linear system with by the vector $x_n \in K_n$ that minimizes the Euclidean norm of $r_n = Ax_n - b$. The restarted version of the GMRES algorithm that we implemented is presented below (Kelley, 1995):

1. choose $x_0$ and compute $r_0 = b - Ax_0$ and $v_1 = r_0/\|r_0\|$
2. for $j = 1,2, \ldots, m$ do
3.      $h_{ij} = (Av_i, v_i)$ for $i = 1,2, \ldots, j$
4.      $\tilde{v}_{j+1} = Av_j - \sum_{i=1}^{j} h_{ij} v_i$
5.      $h_{j+1,j} = \|\tilde{v}_{j+1}\|$
6.      $v_{j+1} = \tilde{v}_{j+1} / h_{j+1,j}$
7. endfor
8. compute $x_m = x_0 + V_m y_m$ where $y_m$ minimizes $\|\beta e_1 - \tilde{H}_m y\|$ for $y \in \mathbb{R}^m$
9. Restart – compute $r_m = b - Ax_m$
10. if $r_m < \varepsilon$ stop
11. else $x_0 = x_m$, $v_1 = r_m/\|r_m\|$ and go to line 2 (for) (restart iterations)

This algorithm contains a level 2 BLAS operation, namely the matrix-vector product $Av_i$ that has to be evaluated at each iteration and level 1 BLAS operations – vector updates inside the *for* loop. The least squares problem (8) can be solved maintaining a QR factorization of H and it requires O(mN) floating point operations (Kelley, 1995). There are several versions of the GMRES algorithms presented in the literature that work at a block level to improve the efficiency or use techniques to parallelize the computations (Chronopoulos, 1986), (Chronopoulos and Kim, 1992), (Chronopoulos and Swanson, 1996), (Chronopoulos and Kucherov, 2010), (Ranjan et al. 2016). In this paper we focused on using GPU CUDA framework to improve the speed of execution of the GMRES algorithm. Obtaining an increase of the speed of execution through parallelization using GPU is a demanding task since the overhead of memory transfers between main memory and device memory is high and one cannot allocate the entire matrices and vectors needed to obtain the solution only on device memory due to the limited amount of memories currently available on GPU cards. A tradeoff between computations done by the CPU and GPU should be achieved to obtain a good speedup.

4. R support for GPU computing

Currently there are several R packages that support GPU/CUDA framework: *gmatrix* (Moris, 2016), *gpuR* (Determan, 2017), *gputools* (Buckner and Wilson, 2013), *cudaBayesreg* (da Silva, 2011), *HiPLARb* (Szeremi, 2012), *HiPLARM* (Nash and Szeremi, 2012) *Rcula* (Morris, 2013). Some of these packages work only with CUDA while others also support OpenCL or OpenACC. While some packages are designed to solve a single type of problems, others are general purpose packages that allow users to implement different types of algorithms. From this latter category we chose *gmatrix*, *gpuR* and *gputools* and implemented the GMRES iterative algorithm for solving linear systems to evaluate the performances of the current R packages that uses the GPU computing model.

*gmatrix* is a general-purpose package that works with only NVIDIA CUDA toolkit. It has *gmatrix*() and *gvector*() methods that allow users to create matrices and vectors on GPU device and all the computations with these matrices/vectors are performed by default on GPU. Data can be transferred between the host memory (RAM) and device memory (GPU) using two simple methods: *h()* and *g()*. This package implements several numerical operations with matrices and vectors on the GPU such as matrix multiplication, addition, subtraction, the Kronecker product, the outer product, comparison operators, logical operators, trigonometric functions, indexing, sorting, random number generation, etc. A matrix or a vector on device memory can be created as simple as:

```
>A <- gmatrix(1:400, 20, 20)
>g_seq <-gseq(1:20)
>b <- as.gvector(g_seq)
```

Performing algebraic operations with such objects is very simple since most of the arithmetic and logical operators (+, -, *, / ,%%, ^, ==, !=, < , > ,<=, >= &, |) are overloaded:

```
>c <- A %*% b
```

In our implementation of the GMRES algorithm using *gmatrix* package we performed only the matrix-vector product on GPU while the rest of the operations are performed by the CPU. The reason why we chose to send to the GPU only the matrix-vector product is that the level 1 operations start to have a speedup > 1 only for very large vectors (N>5e5) according to (Morris, 2016) and due to GPU memory limitations, we used N between 1e3 and 1e4 which is considerably lower than the threshold where the speedup is greater than one.

*gputools* is another general-purpose package that works with CUDA toolkit. Matrices and vectors are created on the host memory using the normal methods and then they are transferred to the device memory where computations took place. After the matrices and vectors are created, the matrix operations dispatched to the GPU were performed using *gpuMatMult()*. Again, we implemented only level 2 operations on GPU because level 1 operations would not be efficient to send to the GPU due to the high overhead incurred by the memory transfers.

Below is a simple example of how a matrix-vector product can be send for execution to the GPU:

```
>N <- 200
>A <- matrix(rnorm(N * N), nrow = N)
>B <- matrix(rnorm(N), nrow = N)
>C <- gpuMatMult(A,B)
```

*gpuR* is the third package that we tested, and it works both with CUDA and OpenCL backends. It has *gpuMatrix()* and *gpuVector()* methods that allow users to create objects on the host memory but the computations with these objects are performed on GPU in a transparent way. There are also two other methods *vclMatrix()* and *vclVector()* that creates the respective objects directly into the device memory and work asynchronously. By using the asynchronous mode, R will immediately return to the CPU after calling any operation that involves *vcl* objects. When one tries to access data computed using *vcl* objects, but the actual computations done by the GPU are not ready, R will wait until the computations are finished and the data become available.

Below are examples of how some basic matrix operations look like using gupR:

```
>N <- 200
>A <- matrix(rnorm(N * N), nrow = N)
>B <- matrix(rnorm(N), nrow = N)
>gpuA <- gpuMatrix(A, type = "double")
>gpuB <- gpuMatrix(B, type = "double")
>gpuC <-gpuA %*% gpuB
```

The most commonly used operators are overloaded: %*%, +, -, *, /, crossprod, tcrossprod, etc. to work with GPU objects. For GMRES we implemented all numerical operations on GPU using *vcl* objects and methods: this approach speeds up the computation but put a limit through the available GPU memory.

The experimental setup consists of a computer with:
- Intel core I7 4710HQ processor at 2.5GHz;
- 6 MB level 3 cache, 1 MB level 2 cache and 256 KB level 1 cache;
- 16 GB DDR3 RAM;
- NVIDIA GeForce 840M graphic card (it has a Maxwell architecture);
- 2 GB video RAM with a bandwidth of 16 GB/s;
- 384 shader units clocked at 1029 MHz;
- Ubuntu ver. 16.04;
- CUDA framework ver. 8.0.61;
- R ver. 3.2.3;
- *gmatrix* ver. 0.3, *gputools* ver. 1.1, *gpuR* ver. 1.2.1;

The following table shows the speedup of the GPU implementations compared with a serial algorithm (*gmres*() from *pracma* package). We used matrices with dimensions between 1000 and 10000. The size of the problem was limited by the available amount of the graphics card memory. It can be observed that the speedups are rather modest, only *gpuR* showing a speedup between 3 and 4. Increasing the problem size could lead to a better speedup but the limited amount of memory on the graphics card precluded us to use bigger matrices.

| N | *gmatrix* implementation | *gputools* implementation | *gpuR* implementation |
|---|---|---|---|
| 1000 | 1.06 | 0.75 | 0.99 |
| 2000 | 1.28 | 0.77 | 1.11 |
| 3000 | 1.33 | 0.83 | 1.25 |
| 4000 | 1.33 | 0.96 | 1.67 |
| 5000 | 1.36 | 1.04 | 2.33 |
| 6000 | 1.46 | 1.17 | 2.90 |
| 7000 | 1.71 | 1.25 | 3.21 |
| 8000 | 2.25 | 1.30 | 3.75 |
| 9000 | 2.45 | 1.41 | 4.10 |
| 10000 | 2.95 | 1.58 | 4.25 |

Table 1. Running times for different implementations and different size of the problem

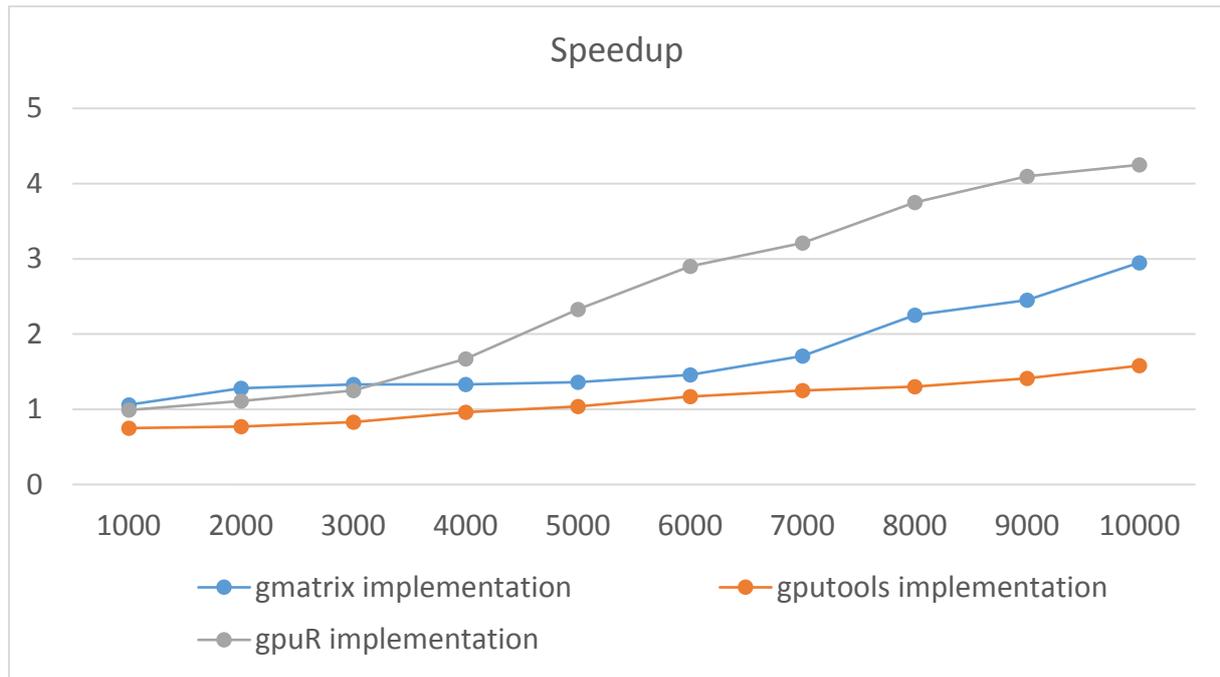

Figure 5. The speed up of the GPU implementations

5. Conclusions

GPU computing becomes an important solution for numerical computing. While C libraries are mature and deliver high speedups, R packages that implements GPU computing are at the beginning. For the three packages that we tested we obtained moderate speedups (~3…4) compared with the serial version of the algorithm. These low speedups are due to the limited amount of device memory that precluded us to use bigger matrices. It is shown in (Morris, 2016) that for level 1 BLAS operations the size of the vector should be greater than 5E5 to obtain speedups greater than 1, but such dimensions are too high for our version of GMRES to fit into the device memory. For commodity computers, where the available amount of graphics card memory is limited the speedups are low and they are comparable with speedups obtained by using a tuned linear algebra library (Oancea et al., 2015). Nevertheless, using GP-GPU together with specialized R packages could be a solution for the future.

**References**

Buckner, M.S., and Wilson, J. (2013), "gputools: A few GPU enabled functions", R package version 0.28, URL  https://CRAN.R-project.org/package=gputools.

Chronopoulos, A.T., and Kucherov, A., (2010), "Block s-step Krylov iterative methods", Numerical Linear Algebra with Applications, Vol. 17, Issue 1, pp. 3-15.

Chronopoulos, A.T, and Swanson, C. D., (1996), "Parallel iterative S-step methods for unsymmetric linear systems", Parallel Computing, Vol. 22, Issue 5, pp. 623-641.

Chronopoulos, A.T., and Kim, S.K., (1992), "Towards Efficient Parallel Implementation of s-step Iterative Methods", Supercomputer, Vol. 47, No. IX-1, pp. 4-17.

Chronopoulos, A.T., (1986), A Class of Parallel Iterative Methods Implemented on Multiprocessors, Ph. D. thesis, Technical Report UIUCDCS-R-86-1267, Department of Computer Science, University of Illinois, Urbana, Illinois, pp. 1-116.

Determan, C. Jr., (2017), "gpuR: GPU Functions for R Objects", R package version 1.2.1., https://CRAN.R-project.org/package=gpuR

Kelley, C.T., (1995), "Iterative Methods for Linear and Nonlinear Equations", Society for Industrial and Applied Mathematics, Philadelphia.

Larsen, E.S., and McAllister, D., (2001), "fast Matrix Multiplies using Graphics Hardware", SC2001 High Performance Computing and Networking Conference, Denver, 10-15 November 2001.

Morris, N. (2016), "Unleashing GPU Power Using R: The gmatrix Package", R package version 0.3., URL https://github.com/njm18/gmatrix

Morris N. (2013), "rcula: An R plugin for matrix factorization and inversion", R package version 0.1, URL https://github.com/njm18/rcula/

Nash, P., and Szeremi, V. (2012), "HiPLARM: High Performance Linear Algebra in R", R package version 0.1, URL https://github.com/cran/HiPLARM

NVIDIA Corporation, (2017), "UDA C Programming Guide".

NVIDIA Corporation, (2007), "NVIDIA CUDA Programming Guide", Version 1.0.

Oancea, B. Andrei, T., and Dragoescu, R.M., (2015), "Accelerating R with high performance linear algebra libraries", Romanian Statistical Review, 63(3), pp. 109-117.

Oancea, B. Andrei, T., and Dragoescu, R.M., (2012a), "GPGPU Computing", Proceedings of the CKS 2012 International Conference, ProUniversitaria, pp. 2026-2035.

Oancea, B. Andrei, T., and Iacob, A.I., (2012b), "Numerical computations in Java with CUDA", 2nd World Conference on Information Technology 2011, published in AWER Procedia Information Technology & Computer Science, 1, pp. 282-285.

Oancea, B., and Andrei, T., (2013), "Developing a High Performance Software Library with MPI and CUDA for Matrix Computations", Computational Methods in Social Sciences, Vol 1, issue 2, pp. 5-10.

OpenCL Specification, Version 1.0, (2008), Khronos Group, Available at: www.khronos.org/registry/OpenCL.

OpenAcc-Standard.org, (2017), The OpenACC Application Programming Interface, Available at: www.openacc.org.

Ranjan, R., Chronopoulos, and A.T., Feng, Y. (2016), "Computational Algorithms for Solving Spectral/hp Stabilized Incompressible Flow Problems", Journal of Mathematics Research, Vol. 8, No. 4, pp. 21-39.

Saad, Y., and M. Schultz, (1986), "GMRES a generalized minimal residual algorithm for solving nonsymmetric linear systems", SIAM Journal on Scientific and Statistical Computing, 7, pp. 856–869.

da Silva, A.F. (2011), "cudaBayesreg: Parallel Implementation of a Bayesian Multilevel Model for fMRI Data Analysis" Journal of Statistical Software, 44(4), pp. 1-24.

Szeremi ,V. (2012), "HiPLARb: High Performance Linear Algebra in R", R package version 0.1.3, URL http://www.hiplar.org/hiplar-b.html

Zhang, D. (2017), "R benchmark for High-Performance Analytics and Computing (II): GPU Packages", available at: http://www.parallelr.com/r-hpac-benchmark-analysis-gpu/